\documentclass[superscriptaddress,twocolumn,showpacs,amsmath,amssymb]{revtex4}
\usepackage[T1]{fontenc}
\usepackage{graphicx}
\usepackage{dcolumn}
\usepackage{bm}
\usepackage{array}

\begin{document}
\title{Estimation of 2p2h effect on Gamow-Teller transition with Second
Tamm-Dancoff-Approximation}

\author{F. Minato}
\affiliation{Nuclear Data Center, Japan Atomic Energy Agency, 
Tokai 319-1195, Japan}

\date{\today}

\begin{abstract}
Two-particle two-hole (2p2h) effect on the Gamow-Teller (GT) transition for neutron-rich nuclei is studied by the second Tamm-Dancoff approximation (STDA) with the Skyrme interaction. Unstable $^{24}$O and $^{34}$Si, and stable $^{48}$Ca nuclei are chosen to study the quenching and fragmentation of the GT strengths.  Correlation of the 2p2h configurations causes $20$\% quenching and downward shift of GT giant resonances (GTGRs). The residual interaction changing relative angular momentum, appeared in the tensor force part, gives a meaningful effect to the GT strength distributions. In this work, $17$ to $26$ \% of the total GT strengths are brought to high energy region above GTGRs. In particular, the tensor force brings strengths to high energy more than $50$ MeV. STDA calculation within a small model space for 2p2h configuration is also performed and experimental data of $^{48}$Ca is reproduced reasonably.
\end{abstract}

\maketitle

\section{Introduction}
The Gamow-Teller (GT) transition is a suitable probe to investigate nuclear 
spin-isospin responses as well as the information on spin dependent channels 
of the nuclear force. It also contributes to matrix elements of the nuclear 
$\beta$-decay and double $\beta$-decay. The former has to do with the time 
scale of rapid neutron-capture process ($r$-process) \cite{Langanke2003} and 
the prediction of decay heat of fission products important for the reprocessing 
of nuclear waste. The latter gives an important insight on the mechanism of 
neutrino mass \cite{Avignone2008}. Low-lying GT state also points out the 
existence of SU(4) ``supermultiplet'' symmetry in nuclei 
\cite{Chunlin2014,Fujita2014}.

It is well known that the total sum of GT strengths follows the model 
independent Ikeda sum rule \cite{Ikeda}, which is defined as 
$S_{-}-S_{+}=3(N-Z)$. Here, $S_{\pm}$ denote the total sum of GT strength 
for $\beta^\pm$ transitions. However, experimentally observed GT strengths 
around GT giant resonances (GTGRs) are only $50-60\%$ of the total value for 
the wide range of nuclei \cite{Gaarde1983,Rapaport1983,Ichimura2006}. The 
leading origin of this damping (so-called $quenching$) mainly lies in the 
coupling with $\Delta$-hole state or the correlation with higher-order 
configurations (see Refs. \cite{Towner1987,Osterfeld1992,Ichimura2006}, and 
references therein). A recent experiment implies that the mixing of 
second-order configuration such as two-particle two-hole (2p2h) state is more 
important than $\Delta$-hole coupling \cite{Sakai2004,Ichimura2006}. The mixing 
of 2p2h states also accounts for the spreading width of GT strengths, which is 
studied by various theoretical approaches \cite{Fiebig1982,Muto1982,Colo1994,Adachi1983,Kuzmin1984,Rijsdijk1993,Nishizaki1988}.

In the higher-order configuration mixing, the tensor force comes to play an important role, because it becomes more effective when the momentum transfer between two particles is high. Bertsch and Hamamoto studied the effect of the tensor force on the total sum of GT strength of $^{90}$Zr including 2p2h states perturbatively and found that it plays a comparable role to the central force \cite{Bertsch1982}. Orlandini et al. obtained the same result in doubly-magic N=Z nuclei by using an energy weighted sum-rule approach \cite{Orlandini1984}. It was also shown that tensor force plays a important role in matrix elements of $\sigma\tau_+$ operator at low energies (see Ref. \cite{Arima1979}, and references therein). Dro\"{z}d\"{z} et al. also studied the effect of the tensor force by second random-phase-approximation (SRPA) \cite{Drozdz1986}, which includes 2p2h states in a similar way to the standard RPA. However, their result showed a moderate effect of the tensor force in contrast to the other works. This difference might be because they used a weaker tensor force than others. However, any of the above works does not consider the self-consistency in their formalisms which is essential to satisfy the Ikeda sum-rule, so that the effects of the tensor force may not be represented correctly. Here, self-consistency means that one uses a same interaction both in the ground state and the residual two-body interaction. In addition, they investigated excitation energy up to at most 50 MeV. Bai et al. reported with a self-consistent RPA \cite{Bai2009} that the tensor force can bring the GT strengths above 50 MeV already in a one-particle one-hole (1p1h) level.

The 2p2h configuration mixing also affects the $\beta$-decay. It is studied for 
light nuclei neighboring the $\beta$-stability line 
\cite{Arima1979,Shimizu1974a}. In case of nuclei far from the $\beta$-stability 
line, several author discussed the $\beta$-decays with the finite rank 
separable approximation (FRSA) \cite{Severyukhin2014} and the particle 
vibration coupling (PVC) \cite{Niu2015}, which consider the 2p2h configuration 
mixing effectively via coupling to phonon states on the top of quasiparticle 
RPA (QRPA). The tensor force induces further modification of the low-lying GT 
state, and it is already studied in 1p1h level \cite{Minato2014,Mustonen2014} 
and FRSA \cite{Severyukhin2014}.

In this paper, we refocus the tensor force effect on the GT transition with the self-consistent Second Tamm-Dancoff approximation (STDA). As target nuclei, unstable $^{24}$O and $^{34}$Si are chosen for future study of the 2p2h effect on $\beta$-decay. We also investigate $^{48}$Ca for comparison with experimental 
data. STDA is, in fact, of a form omitting the ground state correlation part of Second RPA (SRPA). However, it is not so significant in case of the GT transition \cite{Nishizaki1988,Nguyen1997}, so that STDA is suitable for the 
present purpose, because we are able to carry out it within a less computer resource than SRPA. In addition, STDA as well as SRPA is able to include a large model space and respects the Pauli principle which PVC and FRSA cannot take into account properly.

The content of this paper is the following. Section II provides formalism of STDA and discuss the model space to be used in this work. Sec. III discusses our result of GT strength distribution and the quenching. In Sec. IV, the conclusion of present work and future plan are given.


\section{THEORETICAL METHOD}

The numerical calculation of the self-consistent STDA and SRPA had been difficult to be carried out until recently, but is now developed together with the evolution of computer resources and numerical technique, and is used to study monopole, dipole, quadrupole, and octupole transitions up to $^{90}$Zr 
nuclei by realistic interaction \cite{Papakonstantinou2007,Papakonstantinou2010} as well as effective 
interactions such as Skyrme \cite{Gambacurta,Tohyama,Tohyama2012} and Gogny forces \cite{Gambacurta2012}. The present status of the self-consistent SRPA is summarized in Ref. \cite{Papakonstantinou2014}. We therefore illustrate our formalism used in this work briefly in sec. \ref{formalism}, and discuss model spaces for STDA to be considered in sec. \ref{modelspace}.

\subsection{STDA formalism}
\label{formalism}

STDA in this work uses single particle levels obtained by the Skyrme-Hartree-Fock (SHF) method \cite{Skyrme} assuming spherical nuclear shape. We define the ground state of a nucleus as $|\rm{SHF}\rangle$. SHF is solved in the coordinate space with a box boundary condition, $r_{\rm box}$ with a step size $\Delta r=0.1$ fm. We use SGII Skyrme effective interaction for the central and spin-orbit forces \cite{SGII}, and Te1 for the tensor force 
\cite{Te}. Variations of energy density and the spin-orbit potential by adding tensor force can be found in several papers, for example Ref. \cite{Bai2009,Sagawa2014}.

The basic formalism of STDA is same as SRPA formulated in Refs. \cite{Papakonstantinou2010,Gambacurta,Providencia,Yannouleas}. An excited state $|\lambda\rangle$ with respect to the ground state $|0\rangle$ is described by
\begin{equation}
|\lambda;JM\rangle=Q^\dagger_{\lambda;JM}|0\rangle,
\quad
Q_{\lambda;JM}|0\rangle=0
\end{equation}
where the phonon creation operator, $Q^\dagger,$ are defined as
\begin{equation}
Q_{\lambda;JM}^\dagger
=\sum_{mi}X_{mi}^{\lambda;JM}O_{mi}^{JM\dagger}
+\sum_{\substack{m\le n,i\le j\\J_p,J_h}}
\mathcal{X}_{mnijJ_pJ_h}^{\lambda;JM}\mathcal{O}_{mnijJ_pJ_h}^{JM\dagger}.
\end{equation}
The indices $m$ and $n$ denote particle states, while $i$ and $j$ denote hole states. The operators $O_{mi}^{JM\dagger}$ and $\mathcal{O}_{mnijJ_pJ_h}^{JM\dagger}$ create 1p1h and 2p2h states coupled to the angular momentum $J$ and its projection to $z$-axis $M$, respectively \cite{Papakonstantinou2010}. The two particle states and two hole states in the operator $\mathcal{O}$ are coupled to the angular momentum $J_p$ and $J_h$, respectively. Compared with the phonon creation operator of SRPA \cite{Papakonstantinou2010}, the backward amplitudes characterized by $O_{mi}^{JM}$ and $\mathcal{O}_{mnijJ_pJ_h}^{JM}$ are omitted. Instead, the ground state of STDA $|0\rangle$ is exactly identical to $|\rm{SHF}\rangle$.

We define the numbers of 1p1h and 2p2h configurations in a given model space as $N_1$ and $N_2$, respectively. Only $\Delta T_z=\pm1$ configurations are involved and non-charge exchange ($\Delta T_z=0$) and double charge exchange ($\Delta T_z=\pm2$) configurations are decoupled from the phonon creation operator to reduce the dimension of STDA.

The coefficients $X$ and $\mathcal{X}$ are determined by solving the STDA equation,
\begin{equation}
\left(
\begin{tabular}{cc}
$A$ & $\mathcal{A}_{12}$\\
$\mathcal{A}_{21}$ & $\mathcal{A}_{22}$
\end{tabular}
\right)
\left(
\begin{tabular}{c}
$X^\lambda$\\
$\mathcal{X}^\lambda$
\end{tabular}
\right)
=E_\lambda
\left(
\begin{tabular}{c}
$X^\lambda$\\
$\mathcal{X}^\lambda$
\end{tabular}
\right).
\label{tdaequation}
\end{equation}
The $N_1\times N_1$ submatrix $A$ is the standard 1p1h RPA matrix. The submatrices $\mathcal{A}_{12}$ ($N_1\times N_2$ matrix) and $\mathcal{A}_{21}$ ($N_2\times N_1$) describe the coupling between 1p1h and 2p2h 
states, and $\mathcal{A}_{22}$ ($N_2\times N_2$) describes the coupling between 2p2h states. The analytical forms are given in Ref. \cite{Gambacurta}. We calculate the matrix elements in the submatrices self-consistently, namely, the same interaction as the ground state is used. The rearrangement term appearing in matrices $\mathcal{A}$ \cite{Gambacurta2011b} are also included. Since the matrix in the left hand of Eq. \eqref{tdaequation} is symmetric Hermite one, we can obtain eigenvalues, $E_\lambda$, directly by diagonalization, avoiding the imaginary solution problem, emerged in case of SRPA  \cite{Papakonstantinou2010,Gambacurta,Papakonstantinou2014}. Taking advantage of sparse matrices $\mathcal{A}$, Eq. \eqref{tdaequation} is solved by an appropriate numerical method. In this work, the FEAST linear algebra solver \cite{FEAST} is used.

If one omits the coupling between the 2p2h states, $\mathcal{A}_{22}$ becomes diagonal and reads
\begin{equation}
\begin{split}
&[\mathcal{A}_{22}]_{mnij,m'n'i'j'} \\
&=(\epsilon_m+\epsilon_n-\epsilon_i-\epsilon_j)
\delta_{mm'}\delta_{nn'}\delta_{ii'}\delta_{jj'}\chi(m,n)\chi(i,j),
\end{split}
\label{diagonal}
\end{equation}
where the $\epsilon$ are the single particle energies and $\chi(m,n)$ is the antisymmetrizer between $m$ and $n$ states. This prescription is what is called the diagonal approximation, which reduces a computational task considerably. If this approximation works well, it would be useful for a qualitative discussion of GT quenching. We will discuss the validity of the diagonal approximation in Sec. \ref{diagonalsection}.

Transition matrices of an operator $\hat{F}$ is given by
\begin{equation}
\langle\lambda;JM|\hat{F}|0\rangle
=\sum_{mi}X_{mi}^{\lambda;JM}f_{mi}+\sum_{mnij}\mathcal{X}_{mnijJ_pJ_h}^{\lambda;JM}f_{mnij},
\end{equation}
%
where $f_{mi}\equiv\langle m|\hat{F}|i\rangle$ and 
$f_{mnij}\equiv\langle mn|\hat{F}|ij\rangle$.
In case of the GT transition, where 
$\hat{F}=\sum_{mi} \langle m|\vec{\sigma}\tau_\pm|i\rangle a^\dagger_m a_i$, the 
strength function $B_\pm(GT)$ is written as
\begin{equation}
B_\pm^\lambda(GT)=\left|
\sum_{mi}X_{mi}^{\lambda;J} \langle m|| \sigma \tau_{\pm} ||i\rangle 
\right|^2.
\end{equation}
The 2p2h amplitudes $\mathcal{X}$ do not contribute $B_\pm(GT)$ in case of the one body external field as the present case.

Hereafter, ``GT'' denotes $\beta^-$ type transition.

\subsection{Model Space}
\label{modelspace}

Before carrying out the calculation of GT transition, we sought for an appropriate model space. In our calculation, the continuum states are discretized by introducing a box boundary condition. Cutoff energy of unperturbed 1p1h states is fixed to $E_{cut}^{1p1h}=100$ MeV because it is insensitive to results. We also define the cutoff energy of unperturbed 2p2h state as $E_{cut}^{2p2h}$.

\begin{figure}
\includegraphics[width=0.85\linewidth]{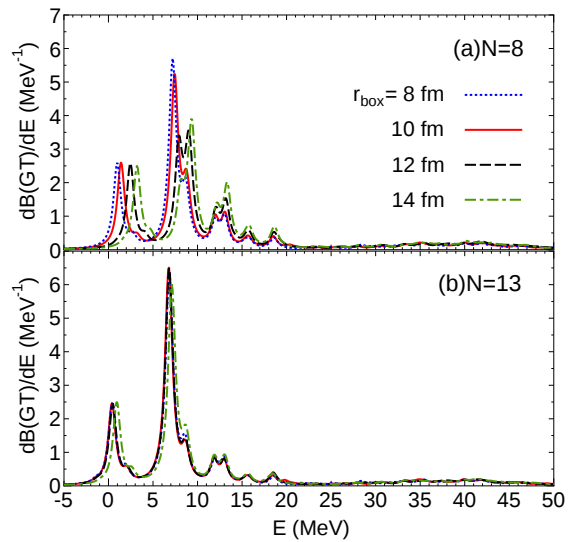}
\caption{(Color online) GT strength distribution of $^{24}$O calculated by STDA(D) with various box boundary conditions from $r_{box}=8$ to $14$ fm. The upper and lower panels show the result for (a) $N=8$ and (b) $N=13$, respectively. The strengths are smeared by the Lorentzian function with a width $1$ MeV. $E_{cut}^{1p1h}$ is set to be $100$ MeV.}
\label{radius}
\end{figure}

\begin{figure}
\includegraphics[width=0.85\linewidth]{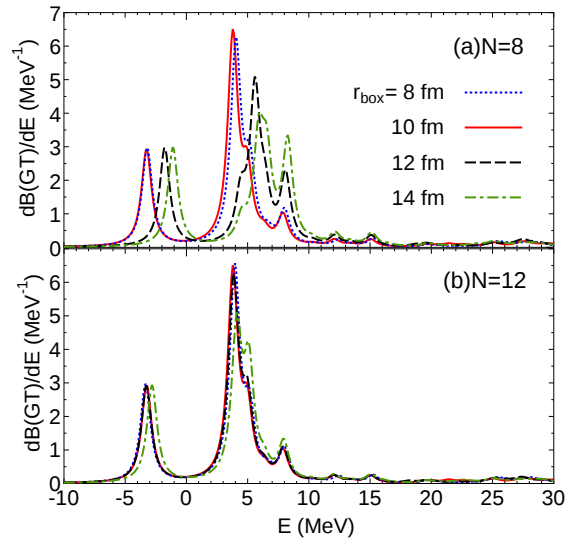}
\caption{(Color online) Same as Fig. \ref{radius}, but for $^{48}$Ca. The upper and lower panels show the result for (a) $N=8$ and (b) $N=12$, respectively.}
\label{radius2}
\end{figure}

First of all, we assessed the boundary box size denoted by $r_{box}$ and a quantum number, $N\equiv 2n+l$, where $n$ and $l$ are the number of node and orbital angular momentum of the single particle wave functions, respectively. 
The test was performed for $^{24}$O and $^{48}$Ca under the condition of the diagonal approximation denoted by STDA(D), by setting $E_{cut}^{2p2h}=100$ MeV. The results are shown in Fig. \ref{radius} and \ref{radius2} for $^{24}$O and $^{48}$Ca, respectively. The horizontal line corresponds to excitation energy of daughter nuclei. We used $Q_\beta$ of the AME mass table \cite{AME}. Let us begin with the upper panel (a) of Fig. \ref{radius}. This is the result of $N=8$ with different $r_{box}$. The strength distributions from $E=-5$ to $20$ MeV do not converge for any $r_{box}$. As we increase $N$, we found that the strength distributions for different $r_{box}$ come close to each other. We finally obtain a good convergence for $N=13$ as shown in the bottom panel (b). The similar result is obtained for $^{48}$Ca. In the upper panel (a) of Fig. \ref{radius2}, the lines for different $r_{box}$ do not converge for $N=8$, while they become almost the same shape for $N=12$ as shown in the bottom panel (b). It means that it is enough to include a large $N$ even in a relatively smaller $r_{box}$ in order to obtain a reasonable convergence, at least within the present condition. We also notice from Fig.\ref{radius} and \ref{radius2} that, in contrast to the energy region from $-5$ to $20$ MeV, the strength distributions at high energies are insensitive to $r_{box}$ and $N$.

\begin{figure}
\includegraphics[width=0.90\linewidth]{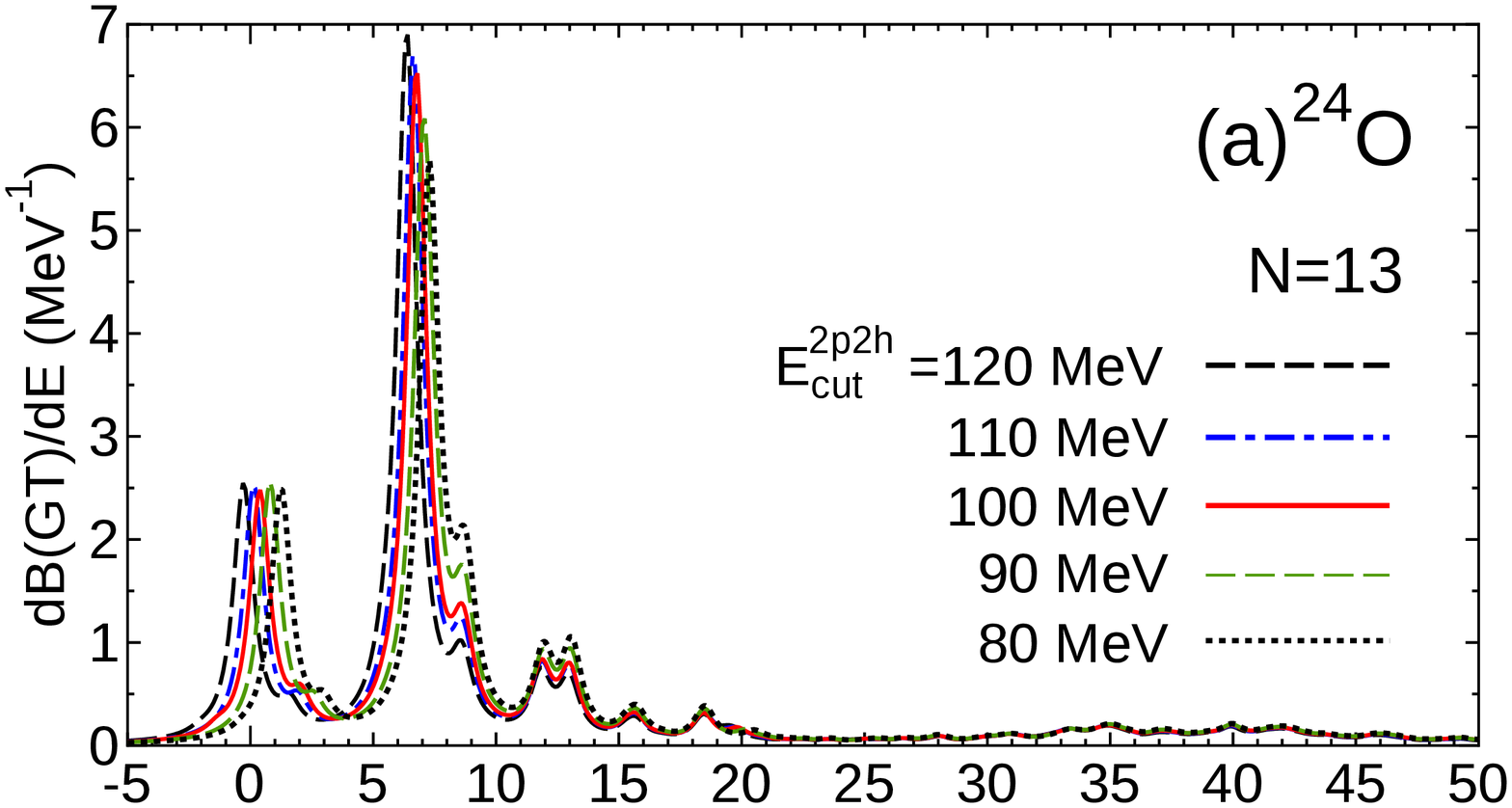}

\includegraphics[width=0.90\linewidth]{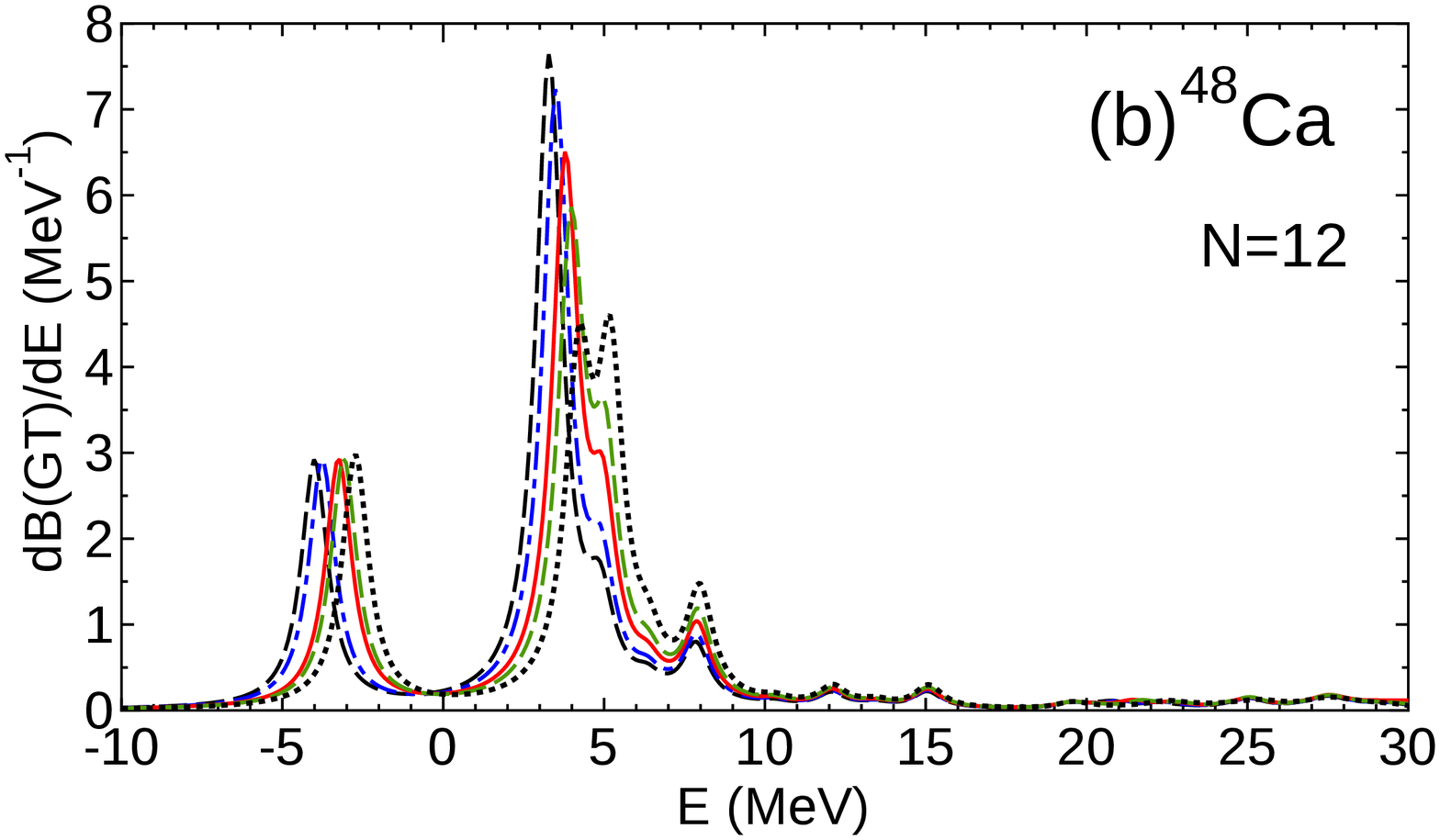}
\caption{(Color online) GT strength distributions of $^{24}$O and $^{48}$Ca calculated by STDA(D) with different cutoff energies of unperturbed 2p2h states denoted by $E_{cut}^{2p2h}$ from $80$ to $120$ MeV.}
\label{cutoff}
\end{figure}

Next, we sought for appropriate $E_{cut}^{2p2h}$. The test calculations is performed altering $E_{cut}^{2p2h}$ from $80$ to $120$ MeV by fixing $N=13$ for $^{24}$O and $N=12$ for $^{48}$Ca. The result is shown in Fig. \ref{cutoff}. The GT resonances up to $10$ MeV show a weak dependence on $E_{cut}^{2p2h}$. Since we use the zero-range interaction in the residual interaction, it is not obvious how large $E_{2p2h}$ should be taken into account. In other words, we have to introduce an appropriate cutoff energy as long as we use the zero-range interaction. However, it would not be so rough to choose a particular $E_{cut}^{2p2h}$ from $80$ to $120$ MeV in order to discuss the average behavior of the GT quenching, considering this weak dependence. We again notice that the GT strength distributions in a high energy region are less sensitive to $E_{cut}^{2p2h}$.

Finally, we adopt $r_{box}=10$ fm, $E_{cut}^{2p2h}=100$ MeV, and $N=13$ for $^{24}$O and $N=12$ for $^{48}$Ca, in this work. The same model space as $^{48}$Ca is adopted for $^{34}$Si. Under this condition, the dimensions $N_1+N_2$ are about $9\times10^4$, $1.4\times10^5$ and $2.5\times10^5$ for $^{24}$O, $^{34}$Si, and $^{48}$Ca, respectively.

\section{RESULTS}
\subsection{2p2h effects on GT distributions}
\label{2p2heffects}
\begin{figure}
\includegraphics[width=0.90\linewidth]{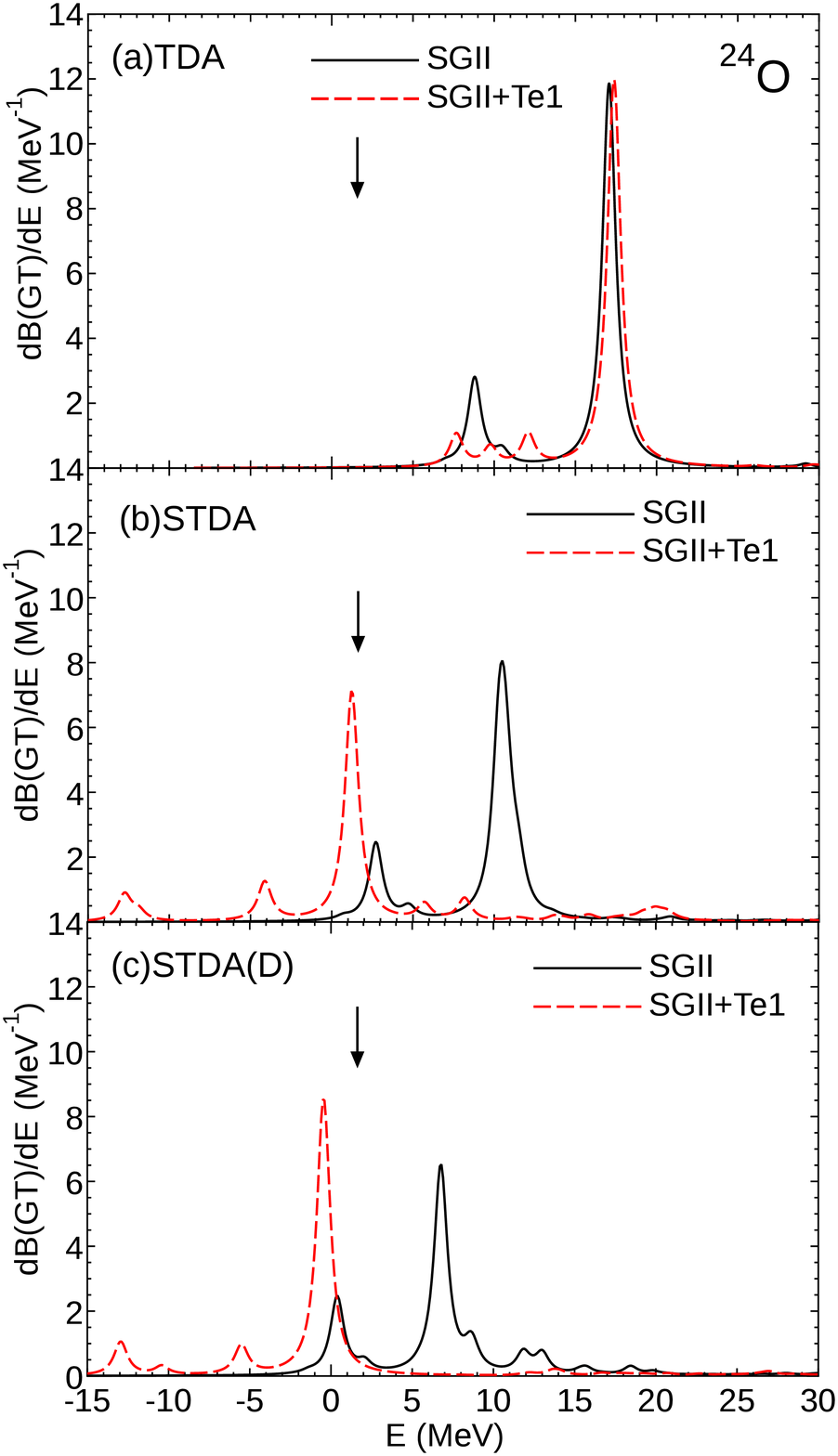}
\caption{(Color online) 
GT strength distribution of $^{24}$O as a function of excitation energy with respect to its daughter nucleus. The upper, middle, and bottom panels are the results for (a)TDA, (b)STDA, and (c)STDA(D), respectively. The solid and dashed lines indicate the results calculated with SGII and SGII+Te1 parameter sets, respectively. The arrow indicates the experimentally observed $1^+$ state \cite{Firestone24}. The strengths are smeared by the Lorentzian function with a width $1$ MeV.}
\label{o24wide}
\end{figure}

We first discuss $^{24}$O. The GT strength distribution as a function of excitation energy of daughter nucleus from $-15$ to $30$ MeV is plotted in Fig. \ref{o24wide}. The GT strengths are smoothed by the Lorentzian function with a width $1$ MeV, which represents the coupling to more complicated states. The position of experimentally observed $1^+$ state ($1.8$ MeV) is indicated by the arrow. The panel (a) illustrates the TDA result. The GTGR appears at about $17$ MeV both for SGII and SGII+Te1. The low-lying resonances can be seen around $9$ MeV for SGII and they are disturbed for SGII+Te1. In case of the STDA shown in the panel (b), the strength distribution of STDA is systematically lower than TDA by about $6$ MeV for SGII and the GTGR appears at about $11$ MeV. This shift can be seen more strongly for SGII+Te1 and several GT peaks appear at negative energies.

The shift of resonances to lower energy region by the coupling with 2p2h states has been commonly observed and discussed in the other SRPA calculations \cite{Papakonstantinou2007,Papakonstantinou2010,Gambacurta,Papakonstantinou2014}. We will discuss this problem later in sec. \ref{downwardsection}.

Let's return to Fig. \ref{o24wide}. The peak height of GTGR for TDA is about $12$. It is quenched to $8$ for STDA as expected from 2p2h configuration mixing. TDA does not show any peaks around $1.8$ MeV where the experimentally observed $1^+$ exists. On the other hand, STDA produces several resonances around this energy by shifting the low-lying GT resonances appearing at about $9$ MeV for TDA. The height of the low-lying resonance is not changed as much as GTGRs. This result is natural because GT states at low energies have less states to be coupled with them than those at high energies. Therefore, the contribution of the 1p1h configuration to this state is still dominant.

\begin{figure}
\includegraphics[width=0.90\linewidth]{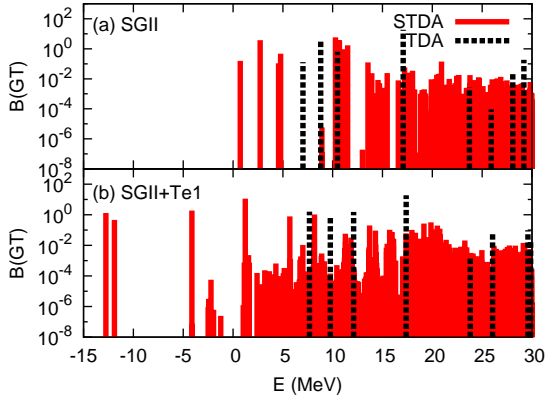}
\caption{(Color online) 
B(GT) of $^{24}$O for (a)SGII and (b)SGII+Te1. The solid and dashed spikes indicate the results of STDA and TDA, respectively.}
\label{log}
\end{figure}

As seen in Fig. \ref{o24wide}, GT strengths are distributed to other states by the 2p2h configuration mixing. To see it more clearly, we plot the discrete GT strength in logarithmic scale in Fig. \ref{log}. The upper panel (a) and bottom panel (b) are the result of SGII and SGII+Te1, respectively. B(GT) of STDA are widely distributed, while that of TDA gives only several peaks around this energy region.

\begin{figure}
\includegraphics[width=0.90\linewidth]{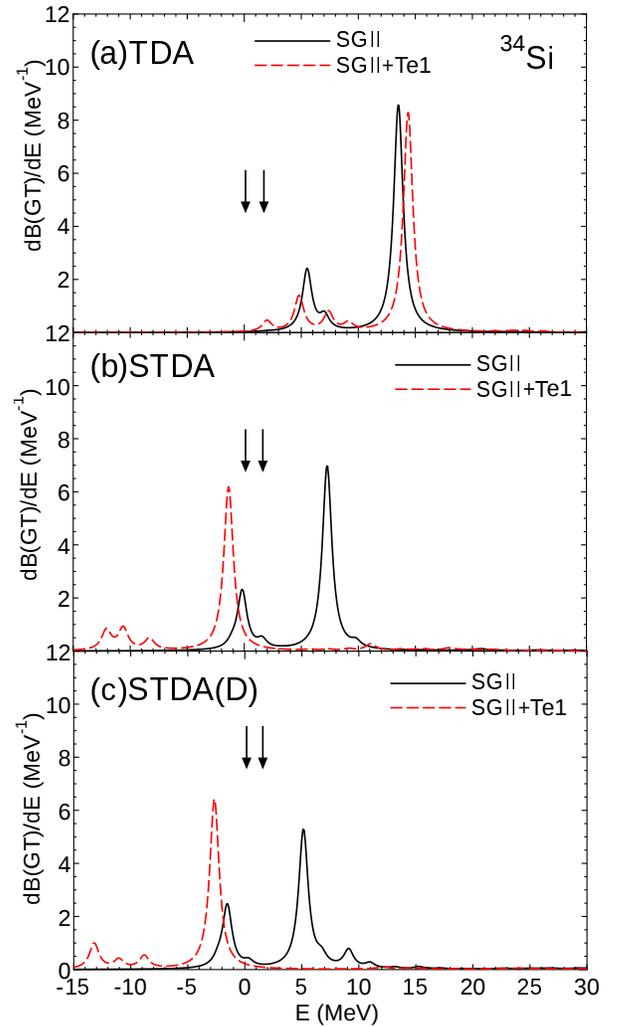}
\caption{(Color online) Same as Fig. \ref{o24wide}, but for $^{34}$Si.}
\label{si34wide}
\end{figure}

Figure \ref{si34wide} shows the GT distribution of $^{34}$Si as a function of excitation energies of its daughter nucleus from $-15$ to $30$ MeV, calculated by (a)TDA, (b)STDA and (c)STDA(D). Two experimentally observed $1^+$ states are denoted by the arrows, one of which is identical to the ground state of the daughter nucleus. The effects of 2p2h correlation and the tensor force are qualitatively same as $^{24}$O. The GT resonances of STDA are lower than TDA by about $7$ MeV for SGII and roughly produce several resonances at which the observed 1$^+$ states exist. SGII+Te1 again shifts the GT resonances downward more strongly in case of STDA and produces the negative resonances. Similar to $^{24}$O, the heights and widths of low-lying resonances are also insensitive to the 2p2h effect.

\begin{figure}
\includegraphics[width=0.90\linewidth]{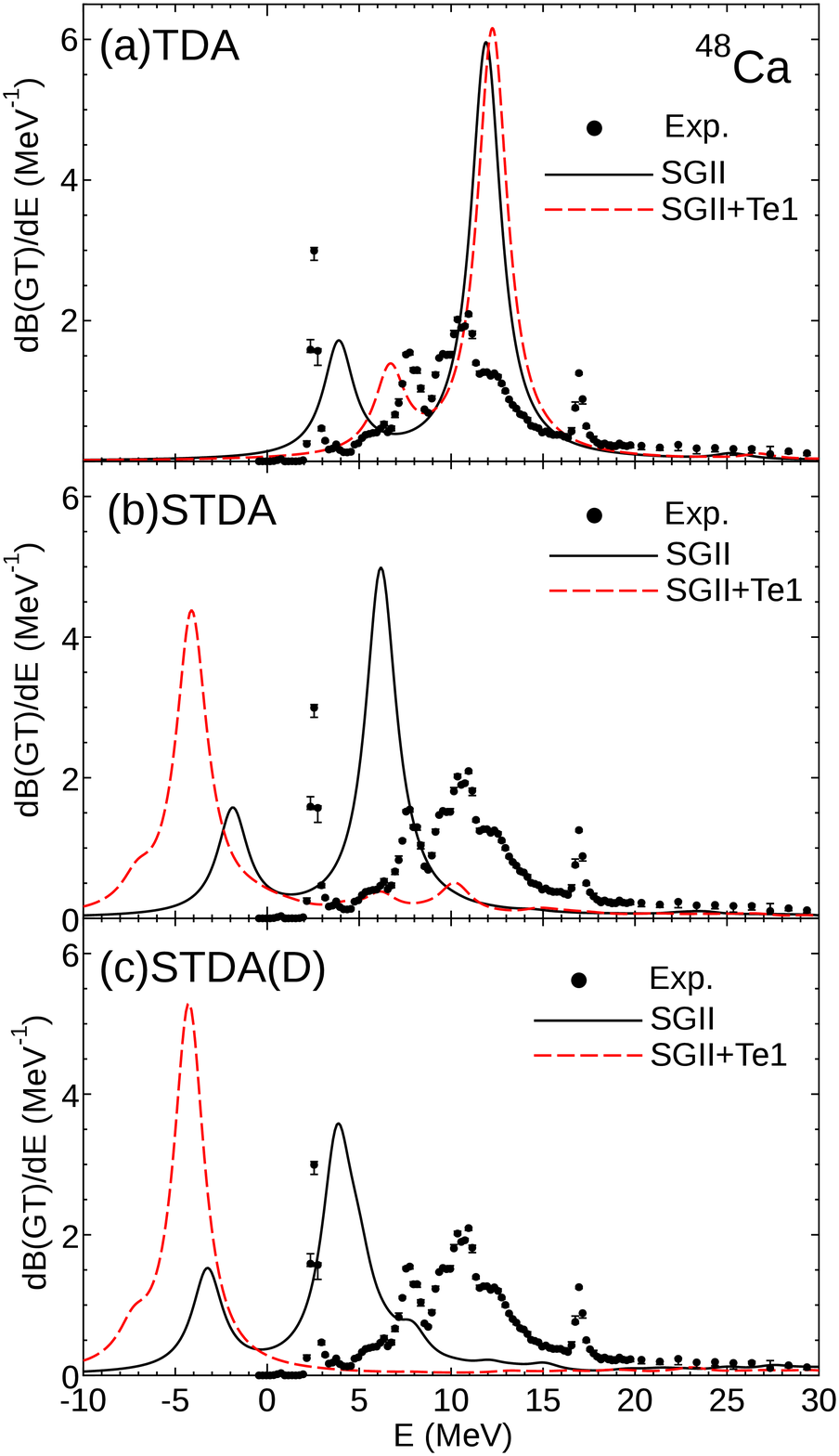}
\caption{(Color online) 
GT strength distribution of $^{48}$Ca as a function of excitation energy with respect to its daughter nucleus. The upper, middle, and bottom panels are the results for (a)TDA, (b)STDA, and (c)STDA(D) respectively. The solid and dashed lines indicate the results calculated with SGII and SGII+Te1 parameter sets, respectively. The experimental data is taken from Ref. \cite{Yako2009}. The strengths are smeared by the Lorentzian function with a width $2$ MeV for comparison with the experimental data as adopted in Ref. \cite{Niu2014}, which is the order of the experimental resolution.}
\label{ca48wide}
\end{figure}

Figure \ref{ca48wide} shows the GT strength distribution of $^{48}$Ca. We plot experimental data measured by Yako et al. \cite{Yako2009} as well. 
TDA roughly reproduces the position of experimental GTGR for SGII and SGII+Te1. The low-lying resonance at $2.5$ MeV is also reproduced fairly if we use SGII. However, widths of the resonances are not reproduced at all due to the lack of coupling with higher-order configurations. STDA produces a slightly wider width, however, the position of GTGR seems rather low and the negative resonances appear as well as $^{24}$O and $^{34}$Si.

\subsection{Tensor force effect and Appearance of negative GT resonance}
\label{downwardsection}

We observed that the GTGRs are significantly moved to lower energies by the tensor force in case of STDA. As a consequence, we obtained the unphysical negative resonances. In particular, the shifts induced by the tensor force was stronger than that by the central force. This would be because the tensor force becomes effective by high momentum transfer between two particles being possible.

In order to investigate the tensor force effect more clearly, we pay attention to the role of the residual interaction. The characteristic of the tensor force is to couple one state formed by two particles with another state different in relative orbital angular momentum $L$ by $2$ as well as $0$. Extracting the relevant angular momentum parts of the interaction matrix elements of tensor forces $V^{tensor}$, emerged in $\mathcal{A}$ of Eq. \eqref{tdaequation}, we obtain
\begin{equation}
\begin{split}
&V_{\mu\nu\mu'\nu'}^{tensor}
\propto \sum_{J'}(-1)^{J'}
\left\{
\begin{tabular}{ccc}
$J'$ & $j_\nu$ & $j_\mu$ \\
$1$  & $j_{\mu'}$ & $j_{\nu'}$
\end{tabular}
\right\} 
\sum_{L,L'}(-1)^{L'}\hat{L}\hat{L'}
\\
&
\times
\left\{
\begin{tabular}{ccc}
$L$ & $1$ & $J'$ \\
$1$  & $L'$ & $2$
\end{tabular}
\right\} 
\left\{
\begin{tabular}{ccc}
$l_\mu$ & $l_\nu$ & $L$\\
$1/2$ & $1/2$ & $1$\\
$j_\mu$ & $j_\nu$ & $J'$\\
\end{tabular}
\right\}
\left\{
\begin{tabular}{ccc}
$l_{\mu'}$ & $l_{\nu'}$ & $L'$\\
$1/2$ & $1/2$ & $1$\\
$j_{\mu'}$ & $j_{\nu'}$ & $J'$\\
\end{tabular}
\right\},
\end{split}
\label{tensor}
\end{equation}
where $\mu$ and $\nu$ can be both particle and hole states, and $j_\mu$ and $l_\mu$ are the total and orbital angular momentums of the state $\mu$ (see \cite{Cao2009} for derivation of Eq. \eqref{tensor}).

We also consider the residual interaction of the central force, where there is a component interacting only among $L=0$ or $2$ states. The relevant angular momentum parts of the interaction matrix elements $V^{center}$ are 
\begin{equation}
\begin{split}
&V_{\mu\nu\mu'\nu'}^{center}
\propto \sum_{L=0,2}
\langle l_\mu||Y_L||l_{\mu'} \rangle \langle l_\nu||Y_L||l_{\nu'} \rangle \\
&\times
\left\{
\begin{tabular}{ccc}
$l_\mu$ & $l_{\mu'}$ & $L$\\
$1/2$ & $1/2$ & $1$\\
$j_\mu$ & $j_{\mu'}$ & $1$\\
\end{tabular}
\right\}
\left\{
\begin{tabular}{ccc}
$l_\nu$ & $l_{\nu'}$ & $L$\\
$1/2$ & $1/2$ & $1$\\
$j_\nu$ & $j_{\nu'}$ & $1$\\
\end{tabular}
\right\}.
\end{split}
\label{central}
\end{equation}
Note that the pairs of orbital angular momentum coupled to $L$ (or $L'$) are different in the center and 
tensor terms. Therefore, we consider them separately. We impose a simple condition that the interaction matrix elements of Eqs. \eqref{tensor} and \eqref{central} are zero unless the following case is satisfied; 
\begin{itemize}
\item case A : $L=0$ for Eq.\eqref{central} (tensor part is excluded)
\item case B : $L=0,2$ for Eq.\eqref{central} (tensor part is excluded)
\item case C : $L=L'$ for Eq.\eqref{tensor} (central part is excluded)
\item case D : $L=L' \& \neq L'$ for Eq.\eqref{tensor} (central part is excluded).
\end{itemize}
Case A retains the relative angular momentum of pairs $[\mu\mu']$ and 
$[\nu\nu']$, and case B allows to exchange it by $2$. Case C also 
retains the relative angular momentum of pairs $[\mu\nu]$ and $[\mu'\nu']$, 
and case D allows it to change.

\begin{table}
\begin{tabular}{c|rrrrr}
     & HF & case A & case B & case C & case D \\
\hline\hline
TDA     & $8.7$ & $15.4$ & $15.6$ & $9.7$ & $8.8$ \\
STDA    & $8.7$ & $11.2$ & $9.4$  & $5.9$ & $-0.9$ \\
STDA(D) & $8.7$ & $10.3$ & $7.9$  & $5.6$ & $-0.7$ \\
\hline
\end{tabular}
\caption{Center of mass of the GT peaks of $^{24}$O defined as 
$E_{cm}$ (see Eq. \eqref{ecom}) in unit of MeV.}
\label{CoM}
\end{table}

\begin{figure}
\includegraphics[width=0.90\linewidth]{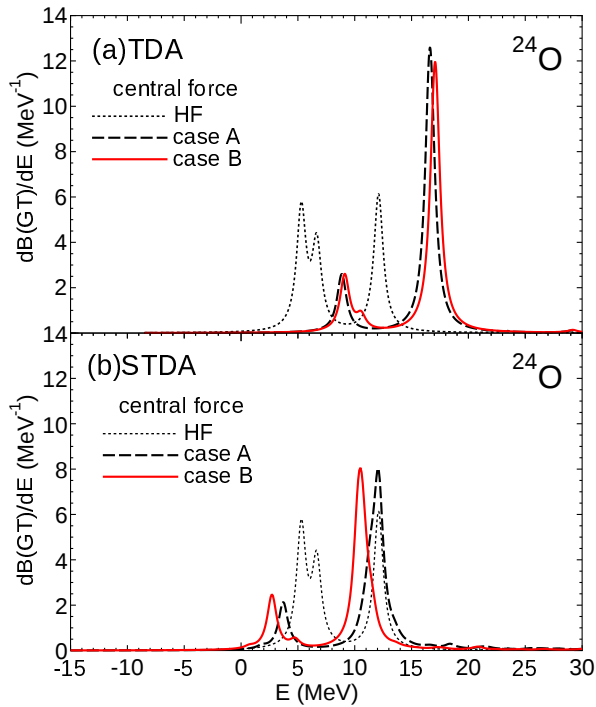}
\caption{(Color online) 
GT strength distribution of $^{24}$O calculated various conditions of the central part of the residual interaction for (a)TDA and (b)STDA(D). The thin solid line indicates the HF solution. The dashed and bold lines are cases A and B, respectively (see the text).}
\label{role1}
\end{figure}

With the above conditions, we calculate the GT strength distribution for $^{24}$O. The center of mass of the GT peaks, defined as 
\begin{equation}
E_{cm}=\frac{\sum_{i\in E_i<30\rm{MeV}} E_iB_-^i(GT)}{\sum_{i\in E_i<30\rm{MeV}} B_-^i(GT)},
\label{ecom}
\end{equation} 
are also calculated and listed in Tab. \ref{CoM}. Figure \ref{role1} shows the results of the central part of the residual interaction, comparing with the Hartree-Fock (HF) result. For TDA shown in the panel (a), the GT resonances for case A appear at higher energies with respect to the HF result and form two sharp peaks. On the other hand, case B gives a minor change and shift $E_{cm}$ upward only by $0.2$ MeV as seen in Tab. \ref{CoM}. Namely, the GT resonances in the 1p1h level is mainly produced for case A and the $L=2$ terms of case B are not of importance. However, we obtain the different result for STDA shown in the panel (b). Case A doesn't push the GT resonances as highly as TDA and case B shifts them to lower energies. 

\begin{figure}
\includegraphics[width=0.90\linewidth]{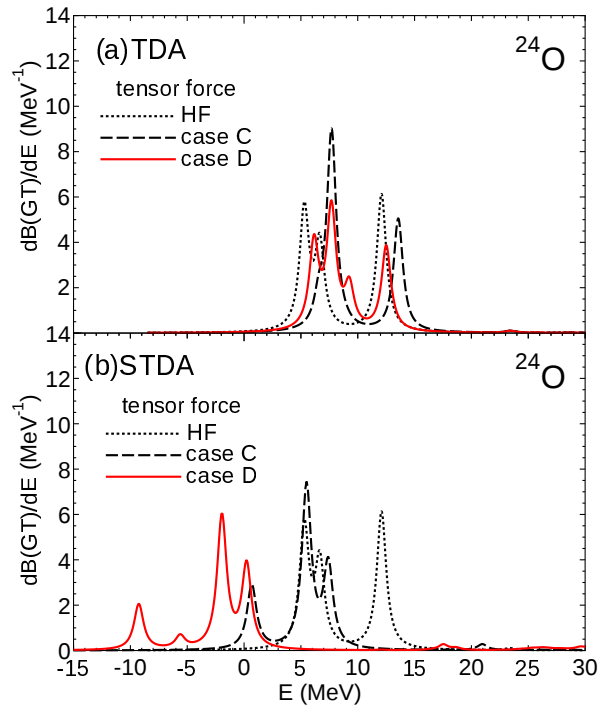}
\caption{(Color online) 
Same as Fig. \ref{role1}, but for the tensor part of the residual interaction. The thin solid line indicates the HF solution. The dashed and bold lines are case C and D, respectively.}
\label{role2}
\end{figure}

Figure \ref{role2} shows the results of the tensor part of the residual interaction. For TDA shown in the panel (a), the GT strength distributions of case C and D aren't significantly different from the HF result. $E_{cm}$ for case C and D are hence close to the one of the HF result, respectively, as shown in Tab. \ref{CoM}. For STDA shown in the panel (b), the resonances of case C move to lower energies with respect to the HF result and $E_{cm}$ is lowered by $2.8$ MeV. Further downward shift can be seen for case D and $E_{cm}$ is $6.8$ MeV lower than the one of case C. In particular, the low-lying GT peak appearing around $1$ MeV for case C shifts downward by as much as $-10$ MeV for case D and appears at around $-9$ MeV. The same result is obtained for other nuclei. 
Therefore, the residual interactions of the tensor force changing the relative orbital angular momentum significantly contributes the downward shift of the GTGRs. The center force part interacting between $L=2$ states has also similar effect, however, its effect is not as strong as the tensor force.


Besides the above discussions, one might think that the use of zero-range interaction contributes the strong shift of GT resonances in case of STDA. It is known that zero-range interactions such as the Skyrme force is not able to properly describe pairing correlation unless an appropriate cutoff energy is introduced \cite{Takahara1994}. Similarly, the use of zero-range interaction in STDA (SRPA) may induce a deviant value in 3 particle-1 hole or 1 particle-3 hole matrix elements appeared in $\mathcal{A}$ of Eq. \eqref{tdaequation}, when high momentum transfers between particles (holes) occur. If one uses a finite-range interaction, a natural cutoff is able to be introduced. However, the downward shift is still commonly observed even when one uses finite-range interactions \cite{Papakonstantinou2010,Gambacurta2012,Papakonstantinou2009}.

It is recently pointed out that the downward shift obtained in the higher-order RPA is substantially attributed to the double counting between the residual interaction and the static response of the ground state \cite{Tselyaev2013,Gambacurta2015}. To overcome this problem, a subtraction method is suggested \cite{Tselyaev2013}. Gambacurta {\it et al.} applied it to SRPA and calculated monopole and quadrupole responses. They showed that too low distributing strength functions obtained by SRPA are pushed up to as high energy as RPA. It is therefore considered that the GT resonances appearing at low energy regions might be also due to the same reason and they are expected to be shifted upward by adopting the subtraction method. 

\subsection{Fragmentation and Quenching of GT strength}
\label{fragmentation}

We observed the fragmented GTGR in STDA and STDA(D) in the previous sections. Some of them are expected to be brought to higher excitation energies. To see them, we plot the GT strength distributions of $^{24}$O, $^{34}$Si, and $^{48}$Ca in high energy regions above $20$ MeV. Because of computational limits, we could not perform STDA at these energy regions so that only STDA(D) was carried out.

\begin{figure}
  \includegraphics[width=0.90\linewidth]{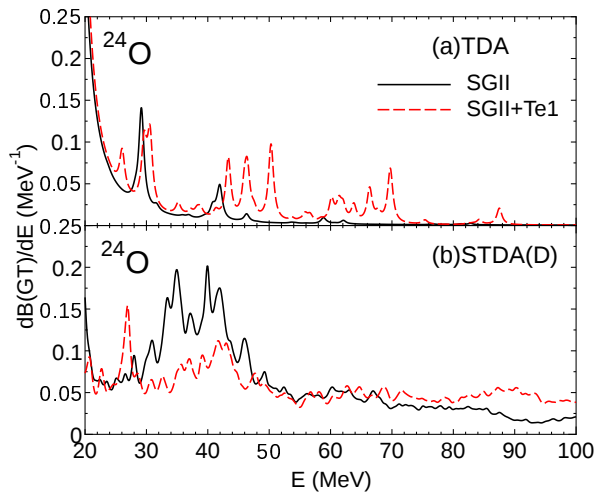}
  \caption{(Color online) GT strength distribution of $^{24}$O as a function of excitation energy with respect to its daughter nucleus. In contrast to Fig. \ref{o24wide}, the horizontal axis is extended up to $100$ MeV. The upper panel and lower panels are (a)TDA and (b)STDA(D) results, respectively. The resonances are smoothed by a Lorentzian function with $1$ MeV width.}
  \label{o24high}
\end{figure}

\begin{figure}
  \includegraphics[width=0.90\linewidth]{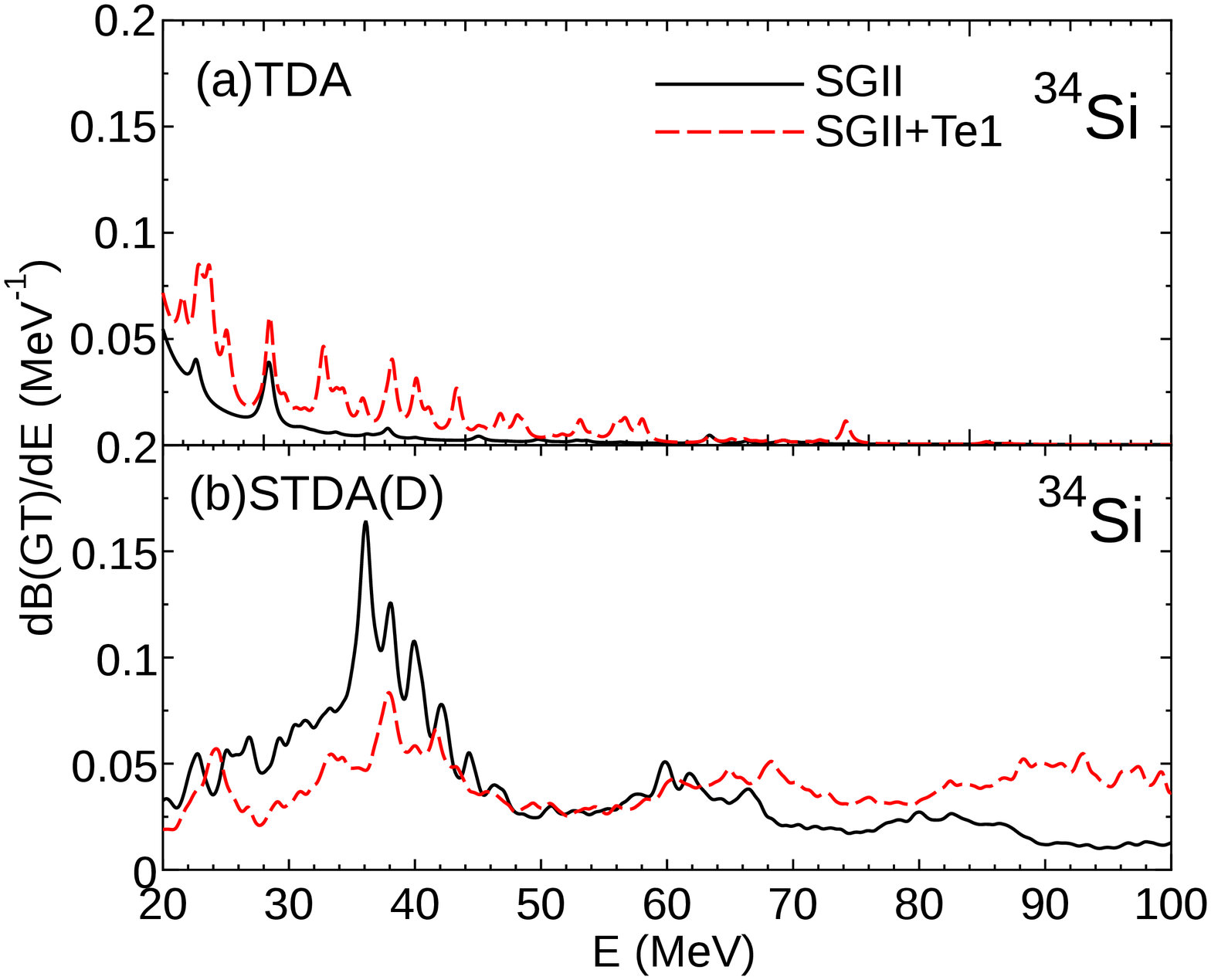}
  \caption{(Color online) Same as Fig. \ref{o24high}, 
    but for $^{34}$Si in the energy region from $20$ to $85$ MeV.}
  \label{si34high}
\end{figure}

\begin{figure}
  \includegraphics[width=0.90\linewidth]{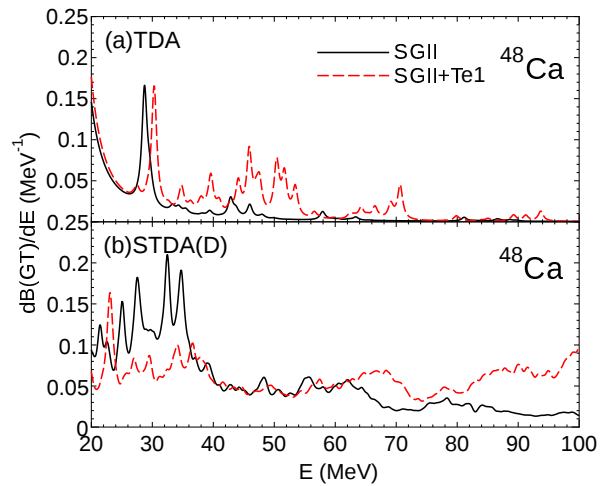}
  \caption{(Color online) Same as Fig. \ref{o24high}, 
    but for $^{48}$Ca in the energy region from $20$ to $80$ MeV.}
  \label{ca48high}
\end{figure}

Let's start by looking at the result of $^{24}$O shown in Fig. \ref{o24high}. For TDA shown in the panel (a), SGII does not give any significant peaks above $50$ MeV, while SGII+Te1 produces a number of peaks there. This result is consistent with that pointed out in Ref. \cite{Bai2009}, in which it is shown that the tensor force brings GT strengths to higher energies even with 1p1h RPA. In case of STDA(D) shown in the panel (b), we can see two marked results. The one is that SGII and SGII+Te1 bring more strengths above $20$ MeV than TDA. The second is that SGII+Te1 significantly enhances the strengths in the energy region above about $60$ MeV, while it suppresses them around $30$-$45$ MeV than SGII. Figures \ref{si34high} and \ref{ca48high} plot the results of $^{34}$Si and $^{48}$Ca at high energy regions, respectively. For both nuclei, STDA(D) brings a number of strengths to higher energy regions and SGII+Te1 further enhances them above approximately $60$ MeV and suppresses around $30$-$40$ MeV as well as $^{24}$O.


To see the enhancement at high energies quantitatively, the sums of the GT strength of $^{24}$O, $^{34}$Si and $^{48}$Ca in percentages of the Ikeda sum rule are listed in Tabs. \ref{sum_o}, \ref{sum_si} and \ref{sum_ca}, respectively. We divide energy region into three parts, which are $E<20$, $20<E<50$, and $50<E$ MeV. We also list the energy weighted sum rule, $m_1$, for numerical check of the present code. Let's first see the result of $^{24}$O. For SGII of TDA, 98.7\% of the Ikeda sum rule is already exhausted in $E<20$. For SGII+Te1 of TDA, it is reduced to 95.2\%. For SGII of STDA(D), the sum of the GT strengths in $E<20$ MeV is meaningfully quenched by 2p2h configuration mixing, being 79.9\%. SGII+Te1 further promotes the quenching and the sum becomes 74.3\%. For STDA, the quenching becomes moderate as compared to STDA(D) and we obtain 86.7\% for SGII and 77.8\% for SGII+Te1 of the Ikeda sum rule in $E<20$ MeV.

Looking at the high energy regions above $20$ MeV, a significant influence of the 2p2h configuration mixing can be seen. For STDA(D), totally 20.4\% and 26.2\% of the Ikeda sum rule are brought to $E>20$ MeV for SGII and SGII+Te1, respectively. In particular, 8.0\% of the Ikeda sum rule is found in $50$ MeV $<E$ for SGII and 17.1\% for SGII+Te1. 

The same results are seen for $^{34}$Si as well. In Tab. \ref{sum_si}, STDA as well as STDA(D) invokes the quenching in the energy region $E<20$ MeV. STDA gives a somewhat weaker quenching than STDA(D). The tensor force further promotes the quenching and the sums in $E<20$ MeV are reduced by 6.2\%. In energy region $20$ MeV $<E$, 17.0\% and 24.9\% of the Ikeda sum rule are found for SGII and SGII+Te1, respectively, and we can still find meaningful GT sums above $50$ MeV.

For $^{48}$Ca, the quenching below $20$ MeV and enhancement above $20$ MeV are qualitatively same as $^{24}$O and $^{34}$Si. The preceding work, the SRPA calculation \cite{Drozdz1986,Nishizaki1988}, estimates that 76\% and 71\% of the Ikeda sum rule are found in about $E<20$ MeV with and without the tensor force, respectively. These values overestimate the experimental data of 58.1\% \cite{Yako2009}. In the present calculation, we obtain 88.5\% and 80.3\% in $E<20$ MeV for SGII and SGII+Te1 of STDA, respectively, which are about 10\% larger than the SRPA calculation. On the other hand, the tensor force effect is closer to the SRPA calculation \cite{Drozdz1986,Nishizaki1988} rather than Bertsch and Hamamoto's result \cite{Bertsch1982}. The present result is also close to that obtained in the PVC calculation \cite{Niu2014}, in case of SGII. 

\begin{table}
  \begin{tabular}{cc|cccc|c}
    $^{24}$O &      & $E$<$20$ & $20$<$E$<$50$ & $50$<$E$ & Total & $m_1$ \\
    \hline\hline
    TDA     & SGII & 98.7 &    1.4   &   0.0   & 100 & 103 \\
        & SGII+Te1 & 95.2 &    3.4   &   1.7   & 100 & 149 \\
\hline
STDA(D) & SGII     & 79.9 &   12.4   &   8.0   & 100 & 104 \\
        & SGII+Te1 & 74.3 &    9.1   &   17.1  & 101 & 152 \\
\hline
STDA    & SGII     & 86.7 &    -     &    -     &   -  & - \\
        & SGII+Te1 & 77.8 &    -     &    -     &   -  & -\\
\end{tabular}
\caption{Sum of GT strengths of $^{24}$O in percentages of the Ikeda sum rule for different energy regions. Energy weighted sum rule, $m_1$, are also listed.}
\label{sum_o}
\end{table}

\begin{table}
\begin{tabular}{cc|cccc|c}
$^{34}$Si&          & $E$<$20$ & $20$<$E$<$50$ & $50$<$E$ & Total & $m_1$ \\
\hline\hline
TDA     & SGII     & 99.8 &    0.6   &   0.0   & 101 & 129 \\
        & SGII+Te1 & 96.2 &    3.6   &   0.7   & 101 & 159 \\
\hline
STDA(D) & SGII     & 83.5 &   10.0   &   7.0   & 100 & 130 \\
        & SGII+Te1 & 77.3 &    6.8   &  16.5   & 101 & 157 \\
\hline
STDA    & SGII     & 88.2 &    -     &    -     &   -  & - \\
        & SGII+Te1 & 79.8 &    -     &    -     &   -  & - \\
\end{tabular}
\caption{Same as Tab.\ref{sum_o}, but for $^{34}$Si.}
\label{sum_si}
\end{table}

\begin{table}
\begin{tabular}{cc|cccc|c}
$^{48}$Ca&          & $E$<$20$ & $20$<$E$<$50$ & $50$<$E$ & Total & $m_1$ \\
\hline\hline
TDA     & SGII     & 99.0 &     1.7  &   0.0   & 101 & 250 \\
        & SGII+Te1 & 95.8 &     4.2  &   0.1   & 101 & 308 \\
\hline
STDA(D) & SGII     & 83.3 &    11.3  &   6.2   & 101 & 249\\
        & SGII+Te1 & 77.0 &     8.1  &  15.7   & 101 & 303 \\
\hline
STDA    & SGII     & 88.5 &     -    &    -    &    - & -\\
        & SGII+Te1 & 80.3 &     -    &    -    &    - & -\\
\end{tabular}
\caption{Same as Tab.\ref{sum_o}, but for $^{48}$Ca.}
\label{sum_ca}
\end{table}

It is known that non-energy ($m_0$) and energy weighted sum rules ($m_1$) of SRPA are analytically identical to those of RPA \cite{Adachi1988}. It also holds for TDA and STDA. To see it, total GT sum (Total) equivalent to $m_0$ and $m_1$ are listed in Tabs. \ref{sum_o}, \ref{sum_si}, and \ref{sum_ca}. We can confirm that TDA and STDA(D) give close values for all the nuclei.

\subsection{Diagonal Approximation}
\label{diagonalsection}

In this section, we see how the diagonal approximation given in Eq. \eqref{diagonal} works. Returning to Figs. \ref{o24wide}, \ref{si34wide}, and \ref{ca48wide}, we can see the GT resonance positions of STDA(D) are lower than those of STDA for all the nuclei. Except it, three interesting findings can be observed from these figures. The one is that the difference of GT resonance position between STDA and STDA(D) becomes smaller as we go from $^{24}$O, which is the lightest nucleus in this work, to the heaviest nuclei, $^{48}$Ca. The second is that the difference between TDA and STDA(D) is much smaller in case of SGII+Te1 than SGII. It implies that the diagonal matrix elements of the tensor force part would be larger than the off-diagonal ones. We can confirm that it is not a wrong anticipation by looking at $E_{cm}$ listed in Tab. \ref{CoM}. For case B, which considers only the central residual interaction, the difference of $E_{cm}$ between STDA and STDA(D) is $1.5$ MeV. On the other hand, if one considers only the tensor residual interaction, that is, case D shows only $0.2$ MeV difference between STDA and STDA(D). The third finding is that the overall GT resonance shape is not significantly different between STDA and STDA(D), for example, relative positions of low-lying GT resonance and GTGR.

From the above results, we would say that the diagonal approximation is a qualitatively good approximation when one calculates heavier nuclei or the tensor force is included.


\subsection{Smaller model space}

To take into account the 2p2h configurations as effectively as possible, a large model space has been introduced in the previous sections. However, the theoretical results do not necessarily reproduce experimental data, especially of $^{48}$Ca.

Toyama and Nakatsukasa studied giant dipole resonances in the $N=82$ isotones by the SRPA calculation. They used a contact force in the residual interaction and a restricted model space smaller than the present study, and reproduced experimental data reasonably well \cite{Tohyama2012}. PVC \cite{Niu2012,Niu2015,Niu2014,Litvinova2014,Marketin2012} also reproduced experimental data successfully using a restricted model space of phonons coupling to 1p1h states. An approach beyond RPA \cite{Nguyen1997} also adopts a relatively smaller model space and shows a good agreement with experimental data.

\begin{figure}
\includegraphics[width=0.90\linewidth]{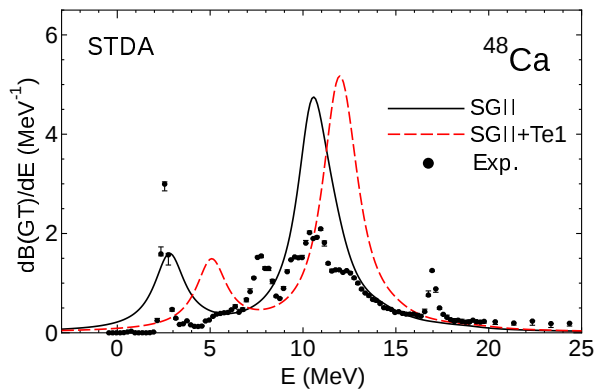}
\caption{(Color online) Same as Fig. \ref{ca48wide} in case of the restricted 
model space (see the text for the detail).}
\label{ca48small}
\end{figure}

We also performed a calculation with a smaller model space. As Toyama and Nakatsukasa did \cite{Tohyama2012}, we consider the single particle levels near the Fermi energy of $1d_{5/2}, 1d_{3/2}, 2s_{1/2}, 1f_{7/2}, 2p_{3/2}, 2p_{1/2}$ and $1f_{5/2}$ orbits of proton and neutron to make the 2p2h states. The neutron and proton orbits are assumed to be fully occupied up to $1f_{7/2}$ and $2s_{1/2}$, respectively. The result is shown in Fig. \ref{ca48small}. The downward shift of GT resonances is inhibited as compared with Fig. \ref{ca48wide}. GTGR appears at around $11$ MeV which is close to the experimental one. The energy of the low-lying resonance is also reasonably reproduced. The obtained GT strength distribution is similar to that obtained in PVC \cite{Niu2014}. The GT resonances of SGII+Te1 locate at higher energy than SGII by about $2$ MeV. This is opposite to that obtained when the larger model space is considered. However, as we increase active single particle levels for 2p2h states, the GT resonances of SGII+Te1 move to lower energies. The sum of GT strengths is also calculated and about 90\% of the Ikeda sum-rule is exhausted for SGII and SGII+Te1 in case of STDA, which is still higher than the experimental value.


\section{Conclusion}
We studied the 2p2h effect on the GT strength distribution for $^{24}$O, $^{34}$Si and $^{48}$Ca. The strength distributions were shifted to lower energy region by the 2p2h configuration mixing. The tensor force also induced further downward shift of the GT resonances. We showed that the tensor residual interactions changing relative orbital angular momentum contributes the strong downward shift. The central residual interaction exchanging relative orbital angular momentum also played a similar role, but its effect was relatively small.

One of the purposes of this paper was to see if we are able to describe the quenching of GT strengths by the self-consistent STDA. We have obtained the results that the sum of GT resonances below $20$ MeV was quenched significantly in case of STDA. The tensor force further promoted the quenching of GT strengths and brought the missing ones to higher excitation energies.  In spite of these results, however, we couldn't reproduce the experimental data of $^{48}$Ca, and no improvements from the preceding SRPA  \cite{Drozdz1986,Nishizaki1988} was obtained. It was also found that the effect of the tensor force used in this work was similar to the result obtained in the preceding SRPA calculation, which gave a moderate tensor force effect than the perturbation approach \cite{Bertsch1982}.

In the diagonal approximation, the GT resonances appeared at a lower energy region than full STDA systematically and the resonances around GTGR were changed slightly. We found that the diagonal approximation relatively worked well, as we calculated heavier nuclei or included the tensor force.

Following the method of Tohyama and Nakatsukasa \cite{Tohyama2012}, we calculated the GT distribution of $^{48}$Ca with the smaller model space. The obtained result reproduced the GT distribution of experimental data reasonably although the restricted model space of the single particle levels were chosen appropriately. At the moment, use of a rather restricted model space or cutoff of insignificant matrix elements as did in Ref. \cite{Kuzmin1987}, are the only approaches to apply STDA and SRPA to heavier or deformed nuclei, in which much larger computer resource is usually required. The small model space would be therefore helpful to obtain a physical insight of 2p2h effects as Tohyama and Nakatsukasa did \cite{Tohyama2012}.

Finally, we would like to note about the subtracted STDA and SRPA approaches (sec. \ref{downwardsection}). Some of the GT resonances obtained in this work appeared at negative energies, which are unphysical. However, the subtracted methods might have them shift upward and produce GT distributions at as high energy region as TDA(RPA). To check it is one of our future subjects.

\section*{ACKNOWLEDGMENT}
The author thanks P. Papakonstantinou and M. Tohyama for the kind instructions 
of SRPA, and K. Yako for providing experimental data and instructive 
information. He also thanks K. Mizuyama who also gave helpful advices.

\appendix

\end{document}